# Computational Determination of the Electronic Structure for different Graphene Quantum Dot geometries.

**A Leon**[1,2]**, J E Gomez.**[1,2] **and  F Perez**[1]
[1] Grupo de Óptica y espectroscopia, Universidad Pontificia Bolivariana, Medellín, Circular 1ª N 70-01, Colombia.
[2] Facultad de Ingenieria en Nanotecnologia, Universidad Pontificia Bolivariana, Medellín, Circular 1ª N 70-01, Colombia.

Corresponding author: alexander.leonr@upb.edu.co

**Abstract**. The interaction between carbon nanostructures like quantum dots and radiation can generate different effects inside the nanomaterial, with the use of computational methods such effects can be predicted and optimize the material allowing a desired output. In this work, a theoretical model for pristine graphene quantum dots is studied, allowing to explain the shape and size dependence for the electronic properties and how the bandgap can be tuned with the functionalization of the nanostructure at the edges.

## 1. Introduction

Semiconductor quantum dots (QDs) are well known because of their optoelectronic properties [1, 2]. The optical response dependence on their geometry and temperature dependence has been studied in the context of laser technology [3, 4] by using InAs embedded in InGaAs. Perovskite QDs was reported to have good performance as laser active media [5]. From reduction of graphene oxide, highly photoluminescence QDs (GQDs) was reported [6]. From the development of carbon nanostructures and the optimization of their synthesis techniques, it became a matter of how to control size, shape and optical properties as well. Different types of precursors and routes in order to obtain GQDs can be found in the literature [7-9]. Additionally, computational models help to predict the behaviour of QDs as laser materials [10].

The popularity of GQDs relies on its radiation-matter interactions, making it a good candidate not only as for semiconductor devices applications, but also as fluorescence emitters and sensors [11-15]. The use of organic precursors allows the functionalization of the nanostructures to obtain different types of molecules at the edges of the GQD in order not only to modify the bad gap of the material but also to chemically attach to other structures like molecules, tissues, and cells [16-20]. Most of the applications rely on the radiation absorption and emission effects of the GQDs functionalized at the edges with molecules that accept or donate electrons [21], where the ones that are acceptors lower the lowest unoccupied molecular orbital (LUMO) [22] and the donors raise the highest occupied molecular orbital (HOMO) [23]. A wide known applications consists in attaching metallic ions to the nanostructure so it can couple into a target molecule where it can be detected later by the excitation and fluorescence of the GQDs [24]. It has had recent developments on the functionalization of GQDs with structures akin to ions of toxic compounds as a qualitative characterization method, where a 400 nM sensitivity has been achieved and even a lower limit of 0.6 nM [25]. In recent studies, It has been sensed Pb ions and even distinguish between different type of ions and gases like $NO_2$ for GQDs embedded in aerogels [26]. Another research focus is being directed to solar cells in which he goal was to increase the absorption capacity of these solar cells for energy harvesting. GQDs can be manipulated to absorb specific wavelengths and this property can be exploited to optimize the absorption of solar energy [27, 28]. These new developments take advantage of the optoelectronic properties of the GQDs, radiation-matter interaction and semiconductor phenomena, allowing a wide variety of applications.

Having in mind the synthesis and characterization of GQDs for fluorescence applications, computational studies have been performed, and in this work, the electronic structure of different edged-shaped GQD geometries was analysed. The effects of the GQD edge and size effects on the density of states was studied. In addition, the band structure of the GQD was calculated.

## 2. Computational model.

### 2.1. Geometrical considerations

A graphene crystal lattice is composed of two atoms A and B with lattice vectors $\vec{a}_1$ and $\vec{a}_2$ that conform the Bravais lattice as shown in Figure 1 and vectors $\vec{\delta}_i$ denoting the nearest neighbors.

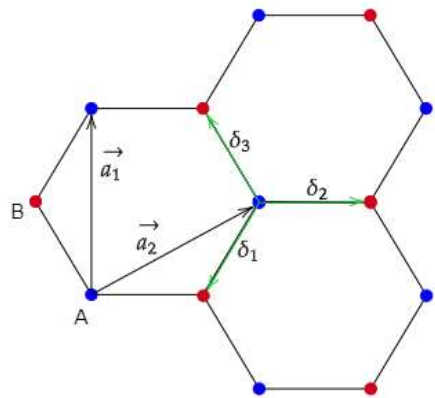

**Figure 1**. Bravais lattice for graphene.

With the use of translational symmetry, a finite size geometry can be constructed. In this research, different zigzag-edged and armchair-edged graphene quantum dots, as showed in Figure 2, were modeled in order to calculate the electronic properties by using the *pybinding* package [29]. This package is based on the *Tight Binding theory*.

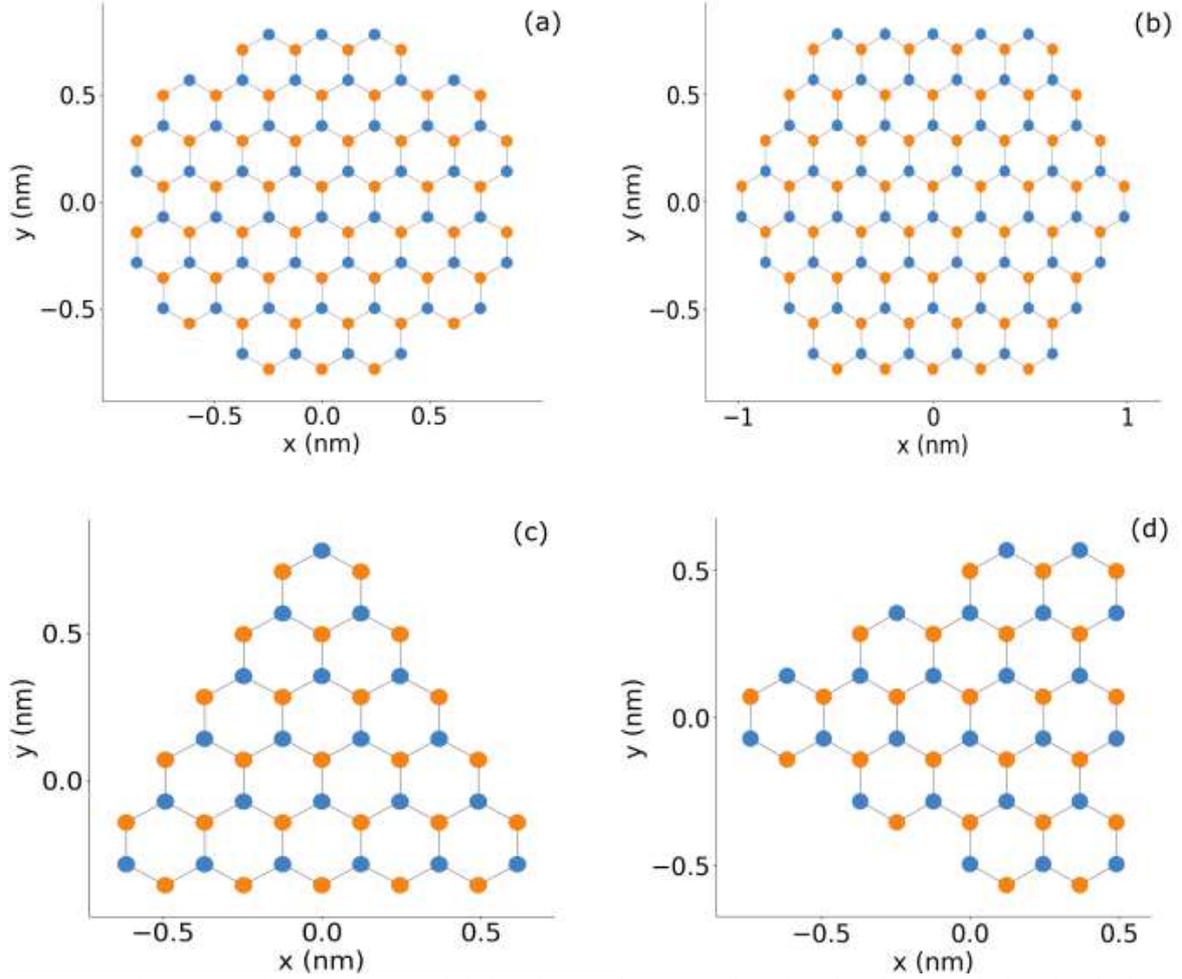

**Figure 2**. Geometries used to model the electronic properties, (a) circular armchair-edged, (b) circular zigzag-edged, (c) triangular zigzag-edged and (d) triangular armchair-edged graphene quantum dots.

*2.2. Tight Binding approach*

A second quantization tight binding model was used to study the electronic structure of the graphene and graphene quantum dots, in this case we consider a system of free, non-interacting fermions given by the following Hamiltonian $\hat{H}$:

$$\hat{H} = \sum_{k,\sigma} \epsilon_k^{free} \hat{c}_{k\sigma}^\dagger \hat{c}_{k\sigma} \qquad (1)$$

where $\sigma$ denotes the spin states $\pm 1/2$ and $\varepsilon_k^{free}$ represents the kinetic energy of a free particle having $k$ wave vector, $\hat{c}^\dagger_{i\sigma}$ and $\hat{c}_{j\sigma}$ are the creation and annihilation operators respectively and the Hamiltonian describes the electrodynamics inside the crystal lattice. The terms after the kinetic energy can be interpreted as a fermion with a given spin $\sigma$ moving over the nearest neighbouring atoms. Exploiting the use of Bloch functions to represent the periodicity of a crystal lattice and Wannier orbitals [30] to represent a superposition of Bloch orbitals, the Hamiltonian can be rearranged as:

$$\hat{H} = \frac{1}{N}\sum_{i,j,\sigma} \sum_k \epsilon_k^{free} e^{ik\cdot(r_i-r_j)} \hat{c}_{i\sigma}^\dagger \hat{c}_{j\sigma} \qquad (2)$$

As it has been said, In Equation (2), the Hamiltonian descibes the electrodynamics of an electron inside the crystal lattice. Now, to include the nearest neighbouring atoms we set a parameter δ that represents the crystal lattice atoms next to a given orbital site, reorganizing the equation:

$$\hat{H}_{tb} = -\frac{t}{2}\sum_{\delta,k,\sigma}\left(e^{ik\cdot\delta} + e^{-ik\cdot\delta}\right)\hat{c}^{\dagger}_{k\sigma}\hat{c}_{k\sigma}, \text{ with } \tilde{t}_{ij} = \frac{1}{N}\sum_k \epsilon_k^{free} e^{ik\cdot(r_i - r_j)} \quad (3)$$

where *t* is known as the hopping parameter, for a graphene structure is known to be 2.7 eV [31]. The Hamiltonian describes the electrodynamics within a given crystal lattice with lattice vector $\vec{\delta}_i$, that consider the nearest neighbouring atoms, as shown in Figure 1.

## 3. Results and discussion.

For the electronic properties, pristine GQDs with the different geometries as shown in Figure 2, ranging from 46, 48, 108 and 144 atoms for both zigzag and armchair edged nanostructures, were build. The first 20 energy levels were calculated for pristine GQDs, and the results are shown in Figures 3 and 4.

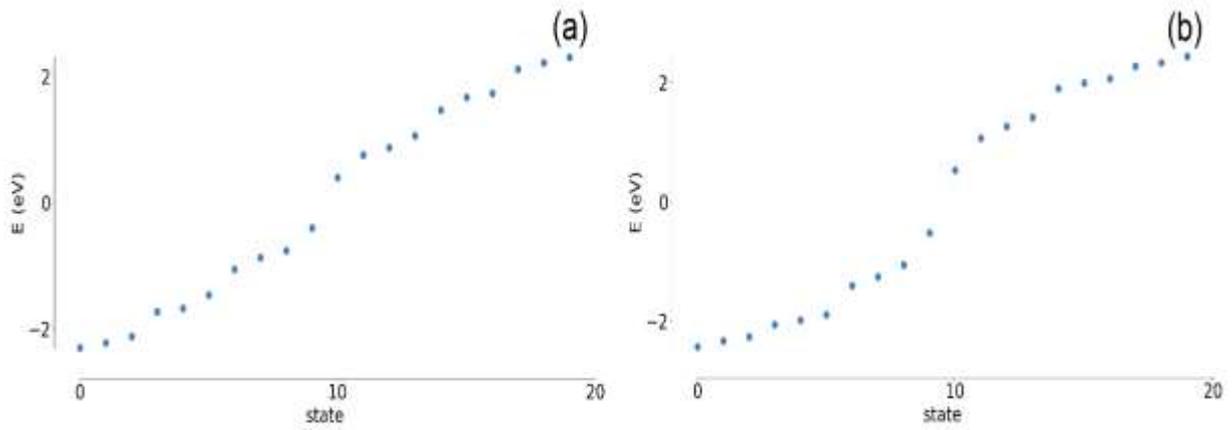

**Figure 3**. First 20 energy levels for (a) zigzag-edged, (b) armchair-edged circular graphene quantum dots.

For the circular GQDs, an approximately 1 eV band gap was found for both armchair and zigzag edged GQDs, hence having a semiconductor behaviour. In Figure 4, the triangular armchair-edged GQD presents an approximately 1.5 eV bandgap following a typical semiconductor behaviour as the circular geometries. On the other hand, the zigzag-edged GQD shows states of energy values around 0 eV, these energies are due to the finite size confinement effects, and the edges bound states in the nanostructure [32].

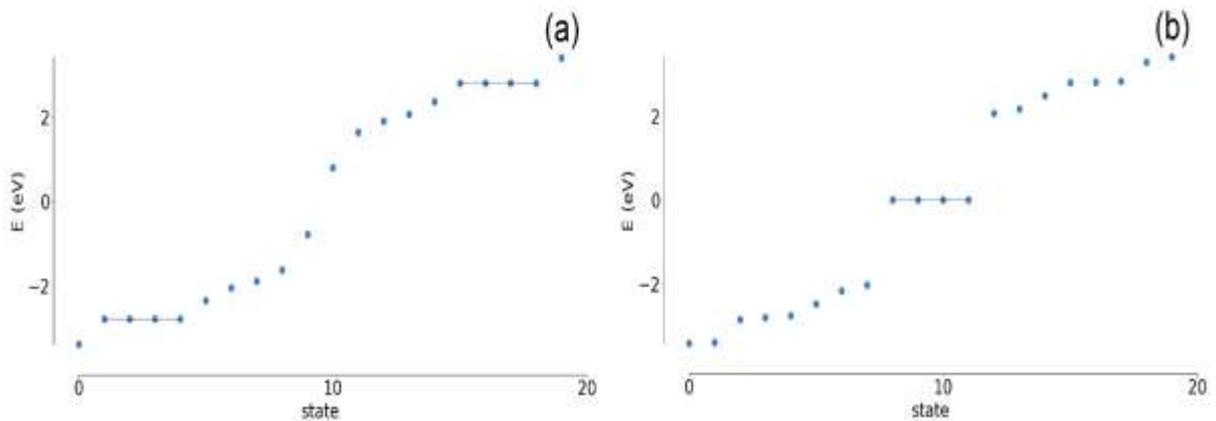

**Figure 4**. First 20 energy levels for (a) armchair-edged, (b) zigzag-edged triangular graphene quantum dots.

Light difference in the band gap structures for the circular armchair-edge and zigzag-edge can be seen from the density of states (DOS) as shown in Figures 5 and 6. The structures of the DOS for the circular armchair-edge and the zigzag-edge, are shown in Figure 5 (a) and (b), respectively.

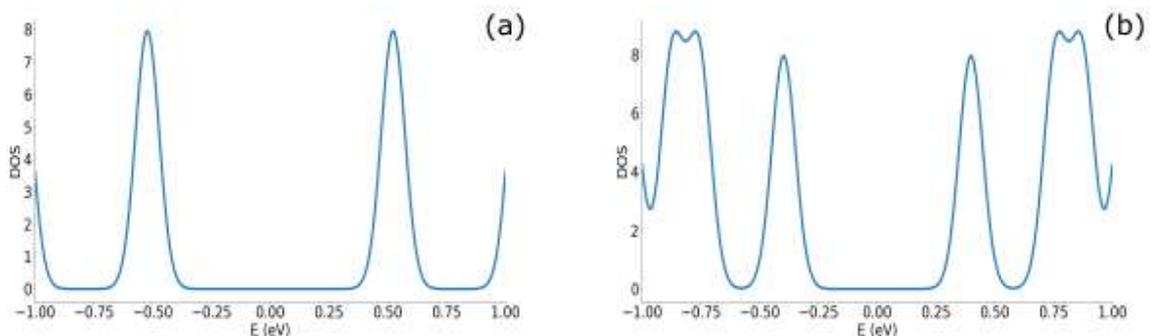

**Figure 5**. Density of states for (a) armchair-edged, (b) zigzag-edged circular graphene quantum dots.

The DOS of the two different circular geometries edges changes, because of the localization of the energy states within the nanostructure. This leads to different optical, electronic, and magnetic interactions and band tuning capabilities reported in the literature [33 - 35]. In the case of triangular geometries, the density of states of the zigzag-edged and the armchair-edged GQD produce them different optical responses thanks to the degenerate levels at the edge states and the spin. This property is observed only in small nanostructures, and in larger size geometries loses, as it is reported in other works [36, 37].

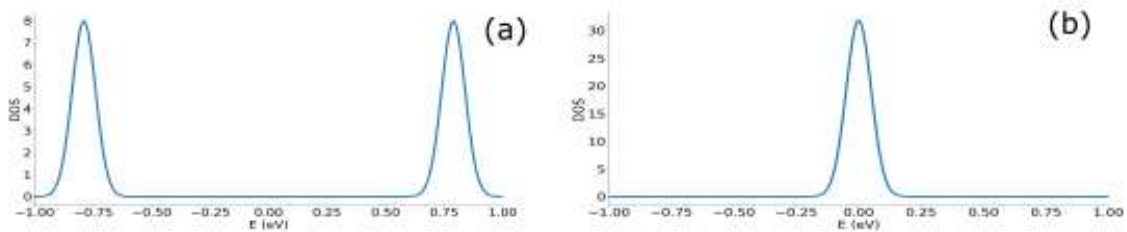

**Figure 6**. Density of states for (a) armchair-edged, (b) zigzag-edged triangular graphene quantum dots.

The band gap and the different DOS are responsible for the semiconductor behaviour of GQD structures being size and edge type dependent for confined-type nanostructures. In fig. 7 and 8, the local density of states (LDOS) for both edge types are shown. It can be seen an edge localization of LDOS on the zigzag-edged geometries.

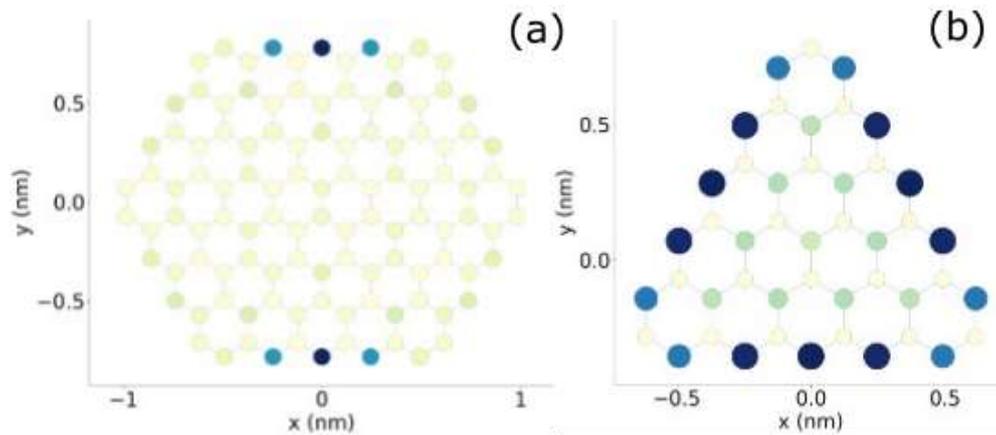

**Figure 7**. Local density of states for (a) circular zigzag-edged and (b) triangular zigzag-edged graphene quantum dots.

The same edge localization effect persists in nanostructures of larger size as seen in Figure 8. The zigzag edges concentrates the LDOS, which could allow the band gap to be fine-tuned when coupled with different functional groups, as has been explored in the literature [38].

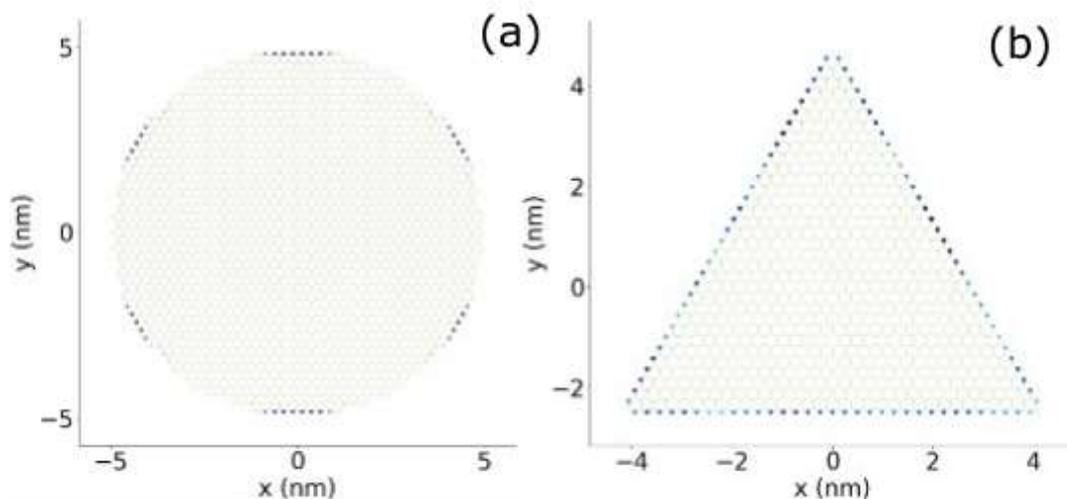

**Figure 8**. Local density of states for (a) 5 nm circular zigzag-edged and (b) 5 nm triangular zigzag-edged graphene quantum dots.

For the case of armchair-edged geometries, the LDOS tends to be localized at the bulk of the nanostructure, as seen in Figure 9. Different sizes for the armchair-edged GQDs were calculated and shows how the LDOS spreads through the nanostructure as its size is increased. Due to this spread of the LDOS into the whole body of the nanostructure, it could exploit the use of dopants and vacancies for band gap manipulation and material optimization [39-41].

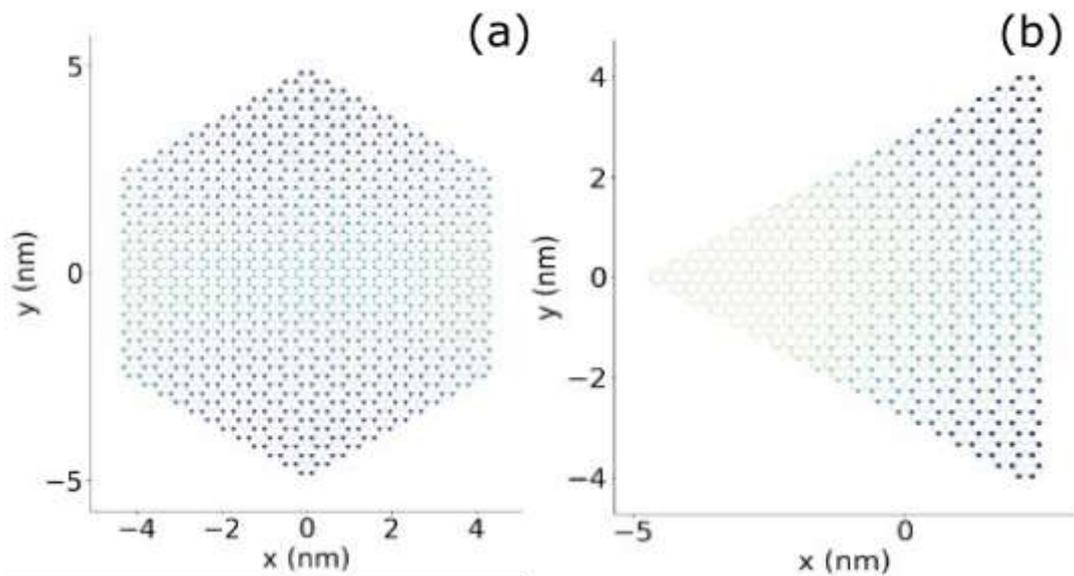

**Figure 9**. Local density of states for (a) 5 nm circular armchair-edged and (b) 5 nm triangular armchair-edged graphene quantum dots

For the semiconductor nanostructures the band strucutre was calculated as shown in Figure 10, where the expected gaps for the 1 eV geometries are present instead of the dirac cones at the k-points that would normally appear on pristine graphene band structures.

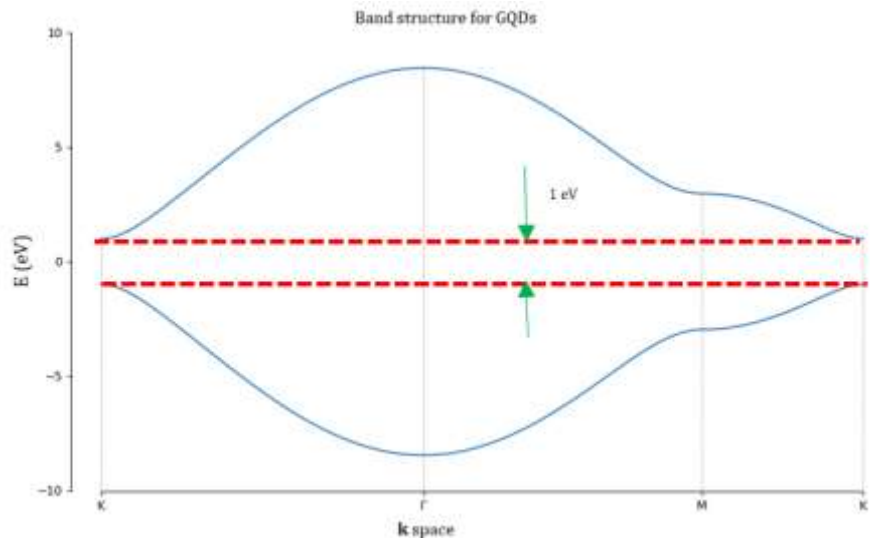

**Figure 10**. Estimated band structure for the semiconductor graphene quantum dots.

## 4. Conclusions.

The electronic properties for different graphene quantum dots were calculated. It was shown how the different edge types in the nanostructures affect the local density of states. This could help in understanding the effects and mechanisms of functionalization mechanisms on the optical properties of GQDs of different sizes and geometries. Computational works on different functionalized nanostructures are needed in order to evaluate how the different chemical groups affect the electronic and optical properties of GQDs.

**Acknowledgments**
The Authors acknowledge Ministry of Science, Technology, and Innovation of Colombia for its financial support to the project by contract number 80740-195-2019, "Optimización de una celda fotoelectrocatalítica para la degradación de contaminantes persistentes basada en modelos fisicoquímicos – Aplicación a colorantes en aguas residuales" Call 808-2018 Proyectos de Ciencia, Tecnología e Innovación y su Contribución a los Retos de País.